\documentstyle[12pt,epsf]{article}
\newcommand{\beq}{\begin{equation}}
\newcommand{\eeq}{\end{equation}}
\newcommand{\beqa}{\begin{eqnarray}}
\newcommand{\eeqa}{\end{eqnarray}}

\begin{document}

\thispagestyle{empty}
\begin{flushright}
SLAC-PUB-7129\\
UCSD-96-05\\
March 1996\\
hep-ph/9603400
\end{flushright}
\vspace*{2cm}
\centerline{{\large\bf Hadro-production of Quarkonia in Fixed Target 
Experiments}\footnote{
Research supported by the Department of Energy under contract 
DE-AC03-76SF00515.}}
\vspace*{1.5cm}
\centerline{{\sc M.~Beneke$^1$} and {\sc I.Z.~Rothstein$^2$}}
\bigskip
\centerline{\sl $^1$Stanford Linear Accelerator Center,}
\centerline{\sl Stanford Univerity, Stanford, CA 94309, U.S.A.}
\vskip0.6truecm
\centerline{\sl $^2$University of California, San Diego,}
\centerline{\sl 9500 Gilman Drive, La Jolla, CA 92093, U.S.A.}

\vspace*{1.5cm}
\centerline{\bf Abstract}
\vspace*{0.2cm}
\noindent 
We analyze charmonium and bottomonium production at fixed 
target experiments. We find that inclusion of color octet 
production channels removes large discrepancies
between experiment and the predictions of the color singlet model
for the total production cross section. Furthermore, including 
octet contributions accounts for the observed direct to total 
$J/\psi$ production ratio. As found earlier for photo-production 
of quarkonia, a fit to fixed target data requires smaller color 
octet matrix elements than those extracted from high-$p_t$ 
production at the Tevatron. We argue that this difference can 
be explained by systematic differences in the velocity expansion 
for collider and fixed-target predictions. While the color 
octet mechanism thus appears to be an essential part
of a satisfactory description of fixed target data, important 
discrepancies remain for the $\chi_{c1}/\chi_{c2}$ production 
ratio and $J/\psi$ ($\psi'$) polarization. These discrepancies, 
as well as, the differences between pion and  proton induced collisions 
emphasize the need for including  higher twist effects in addition to 
the color octet mechanism.
\vspace*{0.5cm}
\noindent 

\vspace*{1.2cm}

\vfill

\newpage
\pagenumbering{arabic}

\section{Introduction}

Quarkonium production has traditionally been calculated in the 
color singlet model (CSM) \cite{SCH94}. Although the model successfully 
describes the production rates for some quarkonium states, 
it has become clear that it fails to provide a theoretically 
and phenomenologically consistent picture of all production 
processes. In hadroproduction of charmonia at fixed target 
energies ($\sqrt{s} < 50\,$ GeV), the ratio  of  the number of $J/\psi$  
produced directly
to those arising from decays of higher 
charmonium states is under-predicted by at least a factor five 
\cite{VAE95}. The $\chi_{c1}$ to $\chi_{c2}$ production ratio 
is far too low, and the observation of essentially unpolarized 
$J/\psi$ and $\psi'$ can not be reproduced. At Tevatron 
collider energies, when fragmentation production dominates, 
the deficit of direct $J/\psi$ and $\psi'$ in the color singlet 
model is even larger. This deficit has been referred to as the 
`$\psi'$-anomaly' 
\cite{BRA94,ROY95}. 

These discrepancies suggest that the color singlet model is 
too restrictive and that other production mechanisms  
are necessitated. Indeed, the CSM requires that the quark-antiquark 
pair that binds into a quarkonium state be produced 
on the time scale $\tau\simeq 1/m_Q$ with the same 
color and angular momentum quantum numbers as the eventually formed 
quarkonium.
Consequently, a hard 
gluon has to be emitted to produce a ${}^3 S_1$ state in the 
CSM and costs one power of $\alpha_s/\pi$. 
Since the time scale for quarkonium formation is of order  
$1/(m_Q v^2)$, where $v$ is the relative quark-antiquark 
velocity in the quarkonium bound state, this suppression 
can be overcome if one allows for the possibility that the 
quark-antiquark pair is in any angular momentum or color 
state when produced on time scales $\tau\simeq 1/m_Q$. 
Subsequent evolution into the physical 
quarkonium state is mediated by emission of soft gluons with momenta 
of order $m_Q v^2$. Since  
the quark-antiquark pair is small in size, the emission of these  
gluons can be analyzed within a multipole expansion.
A rigorous formulation \cite{BOD95} of this picture can be given in 
terms of non-relativistic QCD (NRQCD). Accordingly, the production 
cross section for a quarkonium state $H$ in the process
\begin{equation}
\label{proc}
A + B \longrightarrow H + X,
\end{equation}

\noindent can be written as

\begin{equation}
\label{fact}
\sigma_H = \sum_{i,j}\int\limits_0^1 d x_1 d x_2\,
f_{i/A}(x_1) f_{j/B}(x_2)\,\hat{\sigma}(ij\rightarrow H)\,,
\end{equation}
\begin{equation}
\label{factformula}
\hat{\sigma}(ij\rightarrow H) = \sum_n C^{ij}_{\bar{Q} Q[n]} 
\langle {\cal O}^H_n\rangle\,.
\end{equation}

\noindent Here the first sum extends over all partons in the colliding 
hadrons and $f_{i/A}$ etc. denote the corresponding distribution 
functions. The short-distance ($x\sim 1/m_Q \gg 1/(m_Q v)$) 
coefficients $C^{ij}_{\bar{Q} Q[n]}$ 
describe the production of a quark-antiquark pair 
in a state $n$ and have expansions in $\alpha_s(2 m_Q)$. The 
parameters\footnote{Their precise definition is given in 
Sect.~VI of \cite{BOD95}.} 
$\langle {\cal O}^H_n\rangle$ describe the subsequent 
hadronization of the $Q\bar{Q}$ pair into a jet containing 
the quarkonium $H$ and light hadrons. These matrix elements
 can not be computed 
perturbatively, but their relative importance in powers 
of $v$ can be estimated from the selection rules for multipole 
transitions. 

The color octet picture has led to the most plausible explanation of 
the `$\psi'$-anomaly' and the direct $J/\psi$ production deficit. 
In this picture  
gluons fragment into quark-antiquark pairs in a color-octet 
${}^3 S_1^{(8)}$ state which then hadronizes into a $\psi'$ (or $J/\psi$) 
\cite{BRA95,CAC95,CHO95}. Aside from this striking prediction, the 
color octet mechanism remains largely untested. Its verification 
now requires considering quarkonium production in other processes 
in order to demonstrate the process-independence (universality) of 
the production matrix elements $\langle {\cal O}_n^H\rangle$, which is 
an essential prediction of the factorization formula (\ref{factformula}). 

Direct $J/\psi$ and $\psi'$ production at large $p_t\gg 2 m_Q$ (where 
$m_Q$ denotes the heavy {\em quark} mass) is rather unique in that a 
single term, proportional to $\langle {\cal O}_8^H ({}^3 S_1)\rangle$,  
overwhelmingly dominates the sum (\ref{factformula}). On the other hand,
in quarkonium 
formation at moderate $p_t\sim 2 m_Q$ at colliders and in photo-production 
or fixed target experiments ($p_t\sim 1\,$GeV), 
the signatures of color octet production 
are less dramatic, because they are not as enhanced
by powers of $\pi/\alpha_s$ 
or $p_t^2/m_Q^2$ over the singlet mechanisms. Furthermore,  
theoretical predictions are parameterized by 
more unknown octet matrix elements and  are 
afflicted by larger uncertainties. In particular, there are large 
uncertainties 
due to the increased 
sensitivity to the heavy quark mass close to threshold. (The production 
of a quark-antiquark pair close to threshold is favored by the rise of 
parton densities at small $x$.) 
These facts complicate establishing color 
octet mechanisms precisely in those processes where experimental data is 
most abundant.

Cho and Leibovich \cite{CHO95II} studied direct quarkonium 
production at moderate $p_t$ at the Tevatron collider and were able 
to extract a value for a certain combination of  
unknown parameters $\langle {\cal O}_8^H({}^1 S_0)\rangle$ and 
$\langle {\cal O}_8^H({}^3 P_0)\rangle$ ($H=J/\psi,\psi',\Upsilon(1S),
\Upsilon(2S)$). A first test of universality comes from 
photo-production \cite{CAC96,AMU96,KO96}, where a different combination 
of these two matrix elements becomes important near the elastic 
peak at $z\approx 1$, where $z=p\cdot k_\psi / p \cdot k_\gamma$, and
$p$ is the proton momentum. A fit to photo-production data requires 
much smaller matrix elements than those found in \cite{CHO95II}. 
Taken at face value, this comparison would imply failure of the 
universality assumption underlying the non-relativistic QCD approach. 
However, the extraction from photo-production should be regarded 
with caution since the NRQCD formalism describes 
inclusive quarkonium production only after sufficient smearing in $z$ 
and is not applicable in the exclusive region close to $z=1$, where 
diffractive quarkonium production is important.

In this paper we investigate the universality of the color octet 
quarkonium production matrix elements 
in fixed target hadron collisions and 
re-evaluate the failures of the CSM in fixed target production 
\cite{VAE95} after inclusion of color octet 
mechanisms. Some of the issues involved have already been 
addressed by Tang and V\"anttinen \cite{TAN95} and by Gupta and 
Sridhar \cite{GUP96}, but a complete survey is still missing. We 
also differ from \cite{TAN95} in the treatment of polarized 
quarkonium production and the assessment of the importance of color 
octet contributions and from \cite{GUP96} in the color 
octet short-distance coefficients.

The paper is organized as follows: In Sect.~2 we compile the leading
order color singlet and color octet contributions to the production 
rates for $\psi^\prime,~\chi_J,~J/\psi$ as well as bottomonium. 
In Sect.~3 we present our numerical results for proton and 
pion induced collisions. Sect.~4 is 
devoted to the treatment of polarized quarkonium production. As 
polarization remains one of the cleanest tests of octet quarkonium 
production at large $p_t$ \cite{WIS95,BEN95}, we clarify in detail 
the conflicting treatments of polarized production in \cite{BEN95} and 
\cite{CHO95II}. Sect.~5 is dedicated to a comparison of the extracted 
color-octet matrix elements from fixed target experiments with those 
from photo-production and the Tevatron. We argue that  
kinematical effects and small-$x$ effects can bias the extraction 
of NRQCD matrix elements so that a fit to Tevatron data at large $p_t$ 
requires larger matrix elements than the fit to fixed 
target and photo-production data. The 
final section summarizes our conclusions. 

\section{Quarkonium production cross sections at fixed 
target energies}

\subsection{Cross sections}

We begin with the production cross section for $\psi'$ which does 
not receive contributions from radiative decays of higher charmonium 
states. The $2\to 2$ parton diagrams produce a quark-antiquark pair 
in a color-octet state or $P$-wave singlet state (not relevant 
to $\psi'$) and therefore contribute to $\psi'$ 
production at order $\alpha_s^2 v^7$. (For charmonium $v^2\approx 
0.25 - 0.3$, for bottomonium $v^2\approx 0.08 - 0.1$.) The $2\to3$ 
parton processes contribute to the color singlet processes at order 
$\alpha_s^3 v^3$. Using the notation  in (\ref{fact}):
\begin{eqnarray}
\label{psiprimecross}
\hat{\sigma}(gg\to\psi') &=& \frac{5\pi^3\alpha_s^2}{12 (2 m_c)^3 s}\,
\delta(x_1 x_2-4 m_c^2/s)\left[\langle {\cal O}_8^{\psi'} ({}^1 S_0)
\rangle+\frac{3}{m_c^2} \langle {\cal O}_8^{\psi'} ({}^3 P_0)\rangle
+\frac{4}{5 m_c^2} \langle {\cal O}_8^{\psi'} ({}^3 P_2)\rangle
\right]\nonumber\\[0.0cm]
&&\hspace*{-1.5cm} 
+\,\frac{20\pi^2\alpha_s^3}{81 (2 m_c)^5}\,
\Theta(x_1 x_2-4 m_c^2/s)\,\langle 
{\cal O}_1^{\psi'} ({}^3 S_1)\rangle\,z^2\left[\frac{1-z^2+2 z 
\ln z}{(1-z)^2}+\frac{1-z^2-2z \ln z}{(1+z)^3}\right]\\[0.2cm]
\hat{\sigma}(gq\to\psi') &=& 0\\[0.2cm]
\hat{\sigma}(q\bar{q}\to \psi') &=& \frac{16\pi^3\alpha_s^2}
{27 (2 m_c)^3 s}\,
\delta(x_1 x_2-4 m_c^2/s)\,\langle {\cal O}_8^{\psi'} ({}^3 S_1)
\rangle
\end{eqnarray}

\noindent Here $z\equiv (2 m_c)^2/(s x_1 x_2)$, $\sqrt{s}$ is the 
center-of-mass energy and $\alpha_s$ is normalized at the scale $2 m_c$. 
Corrections to these cross sections are suppressed 
by either $\alpha_s/\pi$ or $v^2$. Note that the 
relativistic corrections 
to the color singlet cross section are substantial in specific 
kinematic regions $z\to 0,1$ \cite{JUN93}. For $\sqrt{s}>15\,$GeV 
these corrections affect the total cross section by less than $50\%$
and decrease 
as the energy is raised \cite{SCH94}. Furthermore, notice that we 
have expressed the short-distance coefficients in terms of the 
charm quark mass, $M_{\psi'}\approx 2 m_c$, rather than the true 
$\psi'$ 
mass. 
Although the difference is formally of higher order in $v^2$, this 
choice is conceptually favored since the short-distance coefficients 
depend only on the physics prior to quarkonium formation. All 
quarkonium specific properties which can affect the cross section, 
such as quarkonium mass differences, 
are hidden in the matrix elements.

The production of $P$-wave quarkonia differs from $S$-waves since 
color singlet and color octet processes enter at the same order in $v^2$ 
as well as $\alpha_s$ in general. An exception is $\chi_{c1}$, which 
can not be produced in $2\to 2$ parton reactions through  
gluon-gluon fusion in a color singlet state. Since at order $\alpha_s^2$, 
the $\chi_{c1}$ would be produced only in a $q\bar{q}$ collision, we 
also include the gluon fusion diagrams at order $\alpha_s^3$, which 
are enhanced by the gluon distribution. We have for $\chi_{c0}$, 

\begin{eqnarray}
\label{chi0cross}
\hat{\sigma}(gg\to\chi_{c0}) &=& \frac{2\pi^3\alpha_s^2}{3 (2 m_c)^3 s}\,
\delta(x_1 x_2-4 m_c^2/s)\frac{1}{m_c^2}
\langle {\cal O}_1^{\chi_{c0}} ({}^3 P_0)
\rangle\\[0.2cm]
\hat{\sigma}(gq\to\chi_{c0}) &=& 0\\[0.2cm]
\hat{\sigma}(q\bar{q}\to \chi_{c0}) &=& \frac{16\pi^3\alpha_s^2}
{27 (2 m_c)^3 s}\,
\delta(x_1 x_2-4 m_c^2/s)\,\langle {\cal O}_8^{\chi_{c0}} ({}^3 S_1)
\rangle\,,
\end{eqnarray}

\noindent for $\chi_{c1}$,

\begin{eqnarray}
\label{chi1cross}
\hat{\sigma}(gg\to\chi_{c1}) &=& \frac{2\pi^2\alpha_s^3}{9 (2 m_c)^5}\,
\Theta(x_1 x_2-4 m_c^2/s)\frac{1}{m_c^2}
\langle {\cal O}_1^{\chi_{c1}} ({}^3 P_1)
\rangle\nonumber\\
&&\hspace*{-1.5cm}
\times\Bigg[\frac{4 z^2\ln z \, (z^8+9 z^7+26 z^6+28 z^5+17 z^4+7 z^3-
40 z^2-4 z-4}{(1+z)^5 (1-z)^4}\nonumber\\
&&\hspace*{-1.5cm}
\,+\frac{z^9+39 z^8+145 z^7+251 z^6+119 z^5-153 z^4-17 z^3-147 z^2-8 z
+10}{3 (1-z)^3 (1+z)^4}\Bigg]
\\[0.2cm]
\hat{\sigma}(gq\to\chi_{c1}) &=& \frac{8\pi^2\alpha_s^3}{81 (2 m_c)^5}\,
\Theta(x_1 x_2-4 m_c^2/s)\frac{1}{m_c^2}
\langle {\cal O}_1^{\chi_{c1}} ({}^3 P_1)
\rangle\left[-z^2\ln z + \frac{4 z^3-9 z+5}{3}\right]\nonumber\\[0.2cm]
\hat{\sigma}(q\bar{q}\to \chi_{c1}) &=& \frac{16\pi^3\alpha_s^2}
{27 (2 m_c)^3 s}\,
\delta(x_1 x_2-4 m_c^2/s)\,\langle {\cal O}_8^{\chi_{c1}} ({}^3 S_1)
\rangle\,,
\end{eqnarray}

\noindent and for $\chi_{c2}$

\begin{eqnarray}
\label{chi2cross}
\hat{\sigma}(gg\to\chi_{c2}) &=& \frac{8\pi^3\alpha_s^2}{45 (2 m_c)^3 s}\,
\delta(x_1 x_2-4 m_c^2/s)\frac{1}{m_c^2}
\langle {\cal O}_1^{\chi_{c2}} ({}^3 P_2)
\rangle\\[0.2cm]
\hat{\sigma}(gq\to\chi_{c2}) &=& 0\\[0.2cm]
\hat{\sigma}(q\bar{q}\to \chi_{c2}) &=& \frac{16\pi^3\alpha_s^2}
{27 (2 m_c)^3 s}\,
\delta(x_1 x_2-4 m_c^2/s)\,\langle {\cal O}_8^{\chi_{c2}} ({}^3 S_1)
\rangle\,.
\end{eqnarray}

\noindent Note that in the NRQCD formalism 
the infrared sensitive contributions to 
the $q\bar{q}$-induced color-singlet process at order $\alpha_s^3$ 
are factorized into the color octet matrix elements 
$\langle {\cal O}_8^{\chi_{cJ}} ({}^3 S_1)\rangle$, so that 
the $q\bar{q}$ reactions at order $\alpha_s^3$ are truly suppressed 
by $\alpha_s$. The production of $P$-wave states through 
octet quark-antiquark 
pairs in a state other than ${}^3 S_1$ is higher order in $v^2$. 

Taking into account indirect production of $J/\psi$ from decays of 
$\psi'$ and $\chi_{cJ}$ states, the $J/\psi$ cross section is given by

\begin{equation}
\label{jpsicross}
\sigma_{J/\psi} = \sigma(J/\psi)_{dir} + \sum_{J=0,1,2} 
\mbox{Br}(\chi_{cJ}\to J/\psi X)\,\sigma_{\chi_{cJ}} + 
\mbox{Br}(\psi'\to J/\psi X)\,\sigma_{\psi'}\,,
\end{equation}

\noindent where `Br' denotes the corresponding branching fraction and 
the direct $J/\psi$ production cross section 
$\sigma(J/\psi)_{dir}$ differs from $\sigma_{\psi'}$ (see 
(\ref{psiprimecross})) only by the replacement of $\psi'$ matrix elements 
with $J/\psi$ matrix elements. Finally, we note that charmonium production 
through $B$ decays is comparatively negligible at fixed target energies.

The $2\to 2$ parton processes contribute only to quarkonium 
production at zero transverse momentum with respect to the beam axis. 
The transverse momentum distribution of $H$ in reaction (\ref{proc})  
is not calculable 
in the $p_t<\Lambda_{QCD}$ region, but the total cross section (which averages 
over 
all $p_t$) is predicted even if the underlying parton process 
is strongly peaked at zero $p_t$.

The transcription of the above formulae to bottomonium production is 
straightforward. Since more bottomonium states exist below the 
open bottom threshold than for the charmonium system, 
a larger chain of cascade decays in the 
bottomonium system must be included. In particular, there is indirect 
evidence from $\Upsilon(3 S)$ production both at the Tevatron 
\cite{PAP95} as well as in fixed target experiments (to be discussed 
below) that there exist yet unobserved $\chi_b(3P)$ states below 
threshold whose decay into lower bottomonium states should also be 
included. Our numerical results do not include indirect contributions 
from potential $D$-wave states below threshold. 

All color singlet cross sections compiled in this section have been 
taken from the review \cite{SCH94}. We have checked that the color 
octet short-distance coefficients agree with those given in 
\cite{CHO95II}, but disagree with those that enter the numerical 
analysis of fixed target data in \cite{GUP96}. 

\subsection{Matrix elements}

The number of independent matrix elements can be reduced by using the 
spin symmetry relations

\begin{eqnarray}
&&\langle {\cal O}^{\chi_{cJ}}_1 ({}^3 P_J) \rangle =
(2 J+1)\,\langle {\cal O}^{\chi_{c0}}_1 ({}^3 P_0)  \rangle
\nonumber\\
&&\langle {\cal O}^{\psi}_8 ({}^3 P_J) \rangle =
(2 J+1)\,\langle {\cal O}^{\psi}_8 ({}^3 P_0)  \rangle
\\
&& \langle {\cal O}^{\chi_{cJ}}_8 ({}^3 S_1) \rangle =
(2 J+1)\,\langle {\cal O}^{\chi_{c0}}_1 ({}^3 S_1)  \rangle
\nonumber
\end{eqnarray}

\noindent and are accurate up to corrections of order $v^2$ ($\psi=J/\psi,
\psi'$ -- identical relations hold for bottomonium). This implies 
that at lowest order in $\alpha_s$, the matrix elements 
$\langle {\cal O}^H_8 ({}^1 S_0) \rangle$ and 
$\langle {\cal O}^H_8 ({}^3 P_0) \rangle$ enter fixed target 
production of $J/\psi$ and $\psi'$ only in the 
combination

\begin{equation}
\label{delta}
\Delta_8(H)\equiv \langle {\cal O}^H_8 ({}^1 S_0) \rangle + 
\frac{7}{m_Q^2}\langle {\cal O}^H_8 ({}^3 P_0) \rangle\,.
\end{equation}

\noindent
Up to corrections in $v^2$, all relevant color singlet production 
matrix elements are related to radial quarkonium wave functions at 
the origin and their derivatives by

\begin{equation}
\label{wave}
\langle {\cal O}^H_1 ({}^3 S_1) \rangle = 
\frac{9}{2\pi} |R(0)|^2
\qquad 
\langle {\cal O}^H_1 ({}^3 P_0) \rangle = 
\frac{9}{2\pi} |R'(0)|^2.
\end{equation}

\noindent We are then left with three non-perturbative parameters for 
the direct production of each $S$-wave quarkonium and two parameters 
for $P$-states.

The values for these parameters, which we will use below, are summarized 
in tables~\ref{tab1} and \ref{tab2}. Many of the octet matrix elements, 
especially for bottomonia, are not established and should 
be viewed as guesses. The numbers given in the tables are motivated as 
follows: All color singlet matrix elements are computed from the
wavefunctions in the Buchm\"uller-Tye potential tabulated in 
\cite{EQ} and using (\ref{wave}). Similar results within $\pm30\%$ 
could be obtained from leptonic and hadronic decays of quarkonia for 
some of the states listed in the tables. The matrix elements 
$\langle {\cal O}_8^H ({}^3 S_1)\rangle$ are taken from the fits to 
Tevatron data in \cite{CHO95II} with the exception of the $3S$ and $3P$ 
bottomonium states. In this case, we have chosen the numbers by (rather 
ad hoc) extrapolation from the $1S$, $2S$ and $1P$, $2P$ states.
\begin{table}[t]
\addtolength{\arraycolsep}{0.2cm}
\renewcommand{\arraystretch}{1.2}
$$
\begin{array}{|c||c|c|c|c|c|}
\hline
\mbox{ME} & J/\psi & \psi' & \Upsilon(1S) & \Upsilon(2S) &
\Upsilon(3S) \\ \hline
\langle {\cal O}^H_1 ({}^3 S_1) \rangle & 
1.16 & 0.76 & 9.28 & 4.63 & 3.54 \\
\langle {\cal O}^H_8 ({}^3 S_1) \rangle & 
6.6\cdot 10^{-3} & 4.6\cdot 10^{-3} & 5.9\cdot 10^{-3} & 
4.1\cdot 10^{-3} & 3.5\cdot 10^{-3} \\
\Delta_8(H)& \mbox{fitted} & \mbox{fitted} & 5.0\cdot 10^{-2} & 
3.0\cdot 10^{-2} & 2.3\cdot 10^{-2} \\
\hline
\end{array}
$$
\caption{\label{tab1} Matrix elements (ME) for the direct 
production of a $S$-wave 
quarkonium $H$. All values in GeV${}^3$.}
\end{table}
\begin{table}[t]
\addtolength{\arraycolsep}{0.2cm}
\renewcommand{\arraystretch}{1.2}
$$
\begin{array}{|c||c|c|c|c|}
\hline
\mbox{ME} & \chi_{c0} & \chi_{b0}(1P) & \chi_{b0}(2P) &
\chi_{b0}(3P) \\ \hline
\langle {\cal O}^H_1 ({}^3 P_0)\rangle/m_Q^2 & 
4.4\cdot 10^{-2} & 8.5\cdot 10^{-2} & 9.9\cdot 10^{-2} & 
0.11 \\
\langle {\cal O}^H_8 ({}^3 S_1) \rangle & 
3.2\cdot 10^{-3} & 0.42 & 0.32 & 0.25 \\
\hline
\end{array}
$$
\caption{\label{tab2} Matrix elements (ME) for the direct 
production of a $P$-wave 
quarkonium $H$. All values in GeV${}^3$.}
\end{table}
The combination of matrix elements $\Delta_8(H)$ turns out to be the 
single most important parameter for direct production of $J/\psi$ and 
$\psi'$. For this reason, we leave it as a parameter to be fitted and later
compared 
with constraints available from Tevatron data. For bottomonia we 
adopt a different strategy since $\Delta_8(H)$ is of no importance 
for the total (direct plus indirect) bottomonium cross section. We therefore 
fixed its value using the results of \cite{CHO95II} together with some 
assumption on the relative size of $\langle {\cal O}_8^H ({}^1 S_0)
\rangle$ and $\langle {\cal O}_8^H ({}^3 P_0)\rangle$ and an ad hoc 
extrapolation for the $3S$ state. Setting $\Delta_8(H)$ to zero for 
bottomonia would change the cross section by a negligible amount.

\section{Results}

Figs.~\ref{prifig} to \ref{upsfig} and table \ref{tab3} summarize our 
results for  the charmonium and bottomonium production cross sections. We 
use the CTEQ3 LO \cite{cteq} parameterization for the 
 parton distributions of  the 
protons and the GRV LO \cite{grv} 
parameterization for pions. The quark masses 
are fixed to be $m_c=1.5\,$GeV and $m_b=4.9\,$GeV, as was done in 
\cite{CHO95II}. The strong coupling is evaluated at the scale 
$\mu=2 m_Q$ ($Q=b,c$) and chosen to coincide with the value 
implied by the parameterization of the parton distributions  
(e.g., $\alpha_s(2 m_c)\approx 0.23$ for CTEQ3 LO). We comment on these 
parameter choices in the discussion below. The experimental data 
have been taken from the compilation in \cite{SCH94} with the addition 
of results from \cite{AKE93} and the $800\,$GeV proton beam at Fermilab 
\cite{SCH95,ALE95}. All data have been 
rescaled to the nuclear dependence $A^{0.92}$ for proton-nucleon 
collisions and $A^{0.87}$ for pion-nucleon collisions. All cross sections 
are given for $x_F>0$ only (i.e. integrated over the forward direction 
in the cms frame where most of the data has been collected).

\subsection{$\psi'$}
\begin{figure}[t]
   \vspace{0cm}
   \epsfysize=8cm
   \epsfxsize=8cm
   \centerline{\epsffile{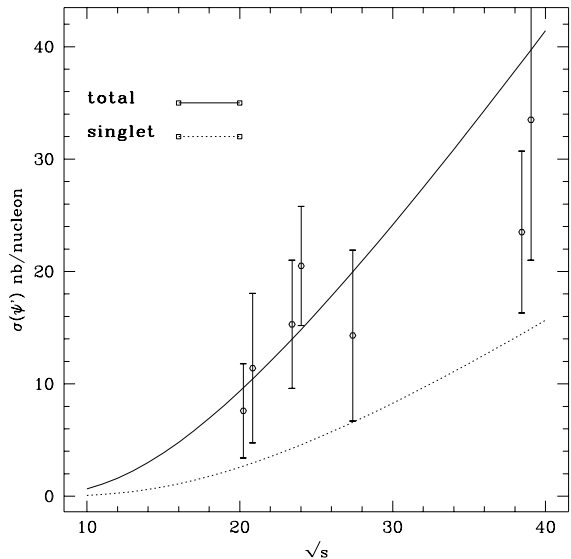}}
   \vspace*{0cm}
\caption{\label{prifig} Total (solid) and singlet only (dotted) 
$\psi^\prime$ production cross section in proton-nucleon collisions 
($x_F>0$ only). The solid line is obtained 
with $\Delta_8(\psi')=5.2\cdot 10^{-3}\,$GeV${}^3$.} 
\end{figure} 

The total $\psi'$ production cross section in proton-nucleus 
collisions is shown in 
Fig.~\ref{prifig}. The color-singlet cross section is seen to be 
about a factor of two below the data and the fit, including color octet 
processes, is obtained with 

\begin{equation}
\Delta_8(\psi') = 5.2\cdot 10^{-3}\,\mbox{GeV}^3\,.
\end{equation}

\noindent The contribution from $\langle {\cal O}^{\psi'}_8({}^3 S_1)
\rangle$ is numerically irrelevant because gluon fusion dominates 
at all cms energies considered here. 
The relative magnitude of singlet and octet contributions 
is consistent with the naive scaling estimate $\pi/\alpha_s\cdot v^4
\approx 1$ (The color singlet cross section acquires an additional 
suppression, because it vanishes close to threshold when 
$4 m_c^2/(x_1 x_2 s)\to 1$). 

It is important to mention that the color singlet prediction has been 
expressed in terms of $2 m_c=3\,$GeV and not the physical quarkonium 
mass. Choosing the quarkonium mass reduces the color singlet cross 
section by a factor of three compared to Fig.~\ref{prifig}, leading to 
an apparent substantial $\psi'$ deficit\footnote{This together with 
a smaller value for the color singlet radial wavefunction could at 
least partially explain the huge discrepancy between the CSM and the  
data that was 
reported in \cite{SCH95}.}. As explained in Sect.~2, choosing 
quark masses is preferred but leads to large normalization uncertainties 
due to the poorly known charm quark mass, which could only be partially 
eliminated if the color singlet wave function were extracted 
from $\psi'$ decays. If, as in open charm production, a small 
charm mass were preferred, the data could be reproduced even without 
a color octet contribution. Although this appears unlikely (see below), 
we conclude that the total $\psi'$ cross section alone does not 
provide convincing evidence for the color octet mechanism.  
If we neglect the color singlet contribution altogether, we obtain 
$\Delta_8(\psi') < 1.0\cdot 10^{-2}\,$GeV${}^3$. This bound is strongly 
dependent on the value of $m_c$. Varying $m_c$ between $1.3\,$GeV and 
$1.7\,$GeV changes the total cross section by roughly a factor of 
eight at $\sqrt{s}=30\,$GeV and even more at smaller $\sqrt{s}$. 
Compared to this normalization uncertainty, the variation with the 
choice of parton distribution and $\alpha_s(\mu)$ is negligible. This 
remark applies to all other charmonium cross sections considered in 
this section.

\subsection{$J/\psi$}

\begin{figure}[t]
   \vspace{0cm}
   \epsfysize=8cm
   \epsfxsize=8cm
   \centerline{\epsffile{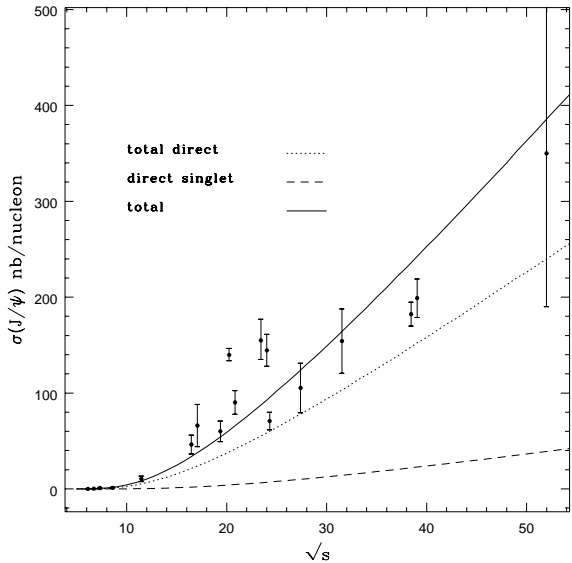}}
   \vspace*{0cm}
\caption{\label{psifig} $J/\psi$ production cross sections in 
proton-nucleon collisions for $x_F>0$. The dotted line is the
direct $J/\psi$ production rate
in the CSM and the dashed line includes  the contribution from  the 
color-octet processes. The total cross section (solid line) includes radiative 
feed-down from the $\chi_{cJ}$ and $\psi'$ states. 
The solid line is obtained 
with $\Delta_8(J/\psi)=3.0\cdot 10^{-2}\,$GeV${}^3$.}
\end{figure}

The $J/\psi$ production cross section in proton-nucleon collisions is 
displayed in Fig.~\ref{psifig}. A reasonable fit is obtained for 

\begin{equation}
\Delta_8(J/\psi)=3.0\cdot 10^{-2}\,\mbox{GeV}^3\,.
\end{equation}

\noindent  We see that  
the color octet mechanism substantially enhances the direct $J/\psi$ 
production cross section compared to the CSM, as shown by the dashed and 
dotted lines in Fig.~\ref{psifig}. The total cross section includes 
feed-down from $\chi_{cJ}$ states which is dominated by the color-singlet 
gluon fusion process. As expected from the cross section in Sect.~2, 
the largest indirect contribution originates from $\chi_{c2}$ states, 
because $\chi_{c1}$ production is suppressed by one power of $\alpha_s$ 
in the gluon fusion channel. The direct $J/\psi$ production fraction 
at $\sqrt{s}=23.7\,$GeV ($E=300\,$GeV) is $63\%$, in excellent agreement 
with the experimental value of $62\%$ \cite{ANT93}. Note that this 
agreement is not a trivial consequence of fitting the color octet 
matrix element $\Delta_8(J/\psi)$ to reproduce the observed total 
cross section since the indirect contribution is dominated by color 
singlet mechanisms and the singlet matrix elements are fixed in terms 
of the wavefunctions of \cite{EQ}.  

\begin{figure}[t]
   \vspace{0cm}
   \epsfysize=8cm
   \epsfxsize=8cm
   \centerline{\epsffile{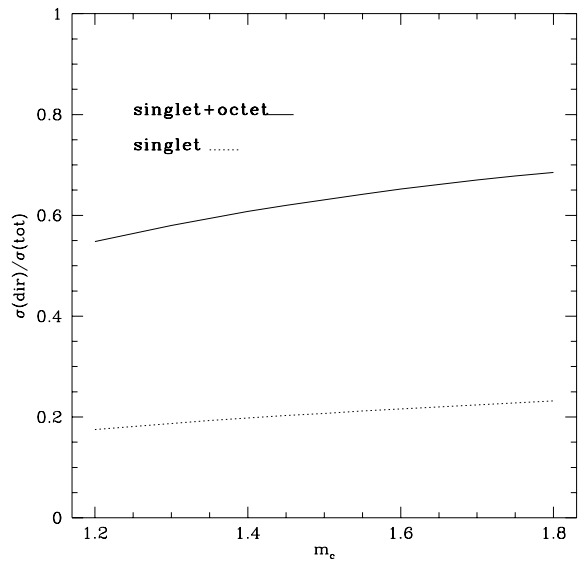}}
   \vspace*{0cm}
\caption{\label{ratfig} Ratio of direct to total $J/\psi$ production 
in proton-nucleon collisions as a function of the charm quark mass in 
the CSM and after inclusion of color octet processes at $E=300\,$GeV. 
The experimental value is $0.62\pm 0.04$.}
\end{figure}

One could ask whether the large sensitivity
to the charm quark mass could be exploited to raise the direct 
production fraction in the CSM, thus obviating the need for octet 
contributions altogether? As shown in Fig.~\ref{ratfig} this is not the 
case, since the charm mass dependence cancels in the direct-to-total 
production ratio. It should be mentioned, that expressing 
all cross sections in terms of the respective quarkonium masses 
increases $\sigma(J/\psi)_{dir}/\sigma_{J/\psi}$, because 
$M_{\chi_{cJ}}>M_{J/\psi}$. However, the total color singlet 
cross section then decreases further and falls short of the data 
by about a factor five. We therefore consider the the 
combination of total $J/\psi$ production cross section and 
direct production ratio as convincing evidence for an 
essential role of color octet mechanisms for direct $J/\psi$ 
production also at fixed target energies.

\begin{table}[b]
\addtolength{\arraycolsep}{0.05cm}
\renewcommand{\arraystretch}{1.1}
$$
\begin{array}{|c||c|c|c||c|c|c|}
\hline
& pN\,\mbox{th.} & pN\,\mbox{CSM} & pN\,\mbox{exp.} & 
\pi^- N\,\mbox{th.} & \pi^- N\,\mbox{CSM} & \pi^- N\,\mbox{exp.}\\ 
\hline
\sigma_{J/\psi} &
90\,\mbox{nb} & 33\,\mbox{nb} & 143\pm 21\,\mbox{nb} & 
98\,\mbox{nb} & 38\,\mbox{nb} & 178\pm 21\,\mbox{nb}\\
\sigma(J/\psi)_{dir}/\sigma_{J/\psi} & 
0.63 & 0.21 & 0.62\pm 0.04 & 
0.64 & 0.24 & 0.56\pm 0.03 \\
\sigma_{\psi'}/\sigma(J/\psi)_{dir} & 
0.25 & 0.67 & 0.21\pm 0.05 & 
0.25 & 0.66 & 0.23\pm 0.05 \\
\chi\mbox{-fraction} & 
0.27 & 0.69 & 0.31\pm 0.04 & 
0.28 & 0.66 & 0.37\pm 0.03 \\
\chi_{c1}/\chi_{c2}\,\mbox{ratio} &
0.15 & 0.08 & - & 
0.13 & 0.11 & 1.4\pm 0.4 \\
\hline
\end{array}
$$
\caption{\label{tab3}
Comparison of quarkonium production 
cross sections in the 
color singlet model (CSM) and  the NRQCD prediction 
(th.) with experiment at $E=300\,$ GeV and $E=185\,$ GeV (last line 
only). The `$\chi$-fraction' is defined by 
$\sum_{J=1,2} \mbox{Br}(\chi_{cJ}\to J/\psi X)\,\sigma_{\chi_{cJ}}/
\sigma_{J/\psi}$. The `$\chi_{c1}/\chi_{c2}$'-ratio is defined by 
$\mbox{Br}(\chi_{c1}\to J/\psi X)\,\sigma_{\chi_{c1}}/
(\mbox{Br}(\chi_{c2}\to J/\psi X)\,\sigma_{\chi_{c2}})$.}
\end{table}

The comparison of theoretical predictions with the E705 experiment 
\cite{ANT93} is summarized in Tab.~\ref{tab3}. Including color octet 
production yields good agreement for direct $J/\Psi$ production, 
as well as the relative contributions from all $\chi_{cJ}$ states 
and $\psi'$. Note that the total cross section from \cite{ANT93} 
is rather large in comparison with other data (see Fig.~\ref{psifig}). 
In the CSM, the direct production cross section of $7\,$nb should 
be compared with the measured $89\,$nb, clearly demonstrating 
the presence of an additional numerically large production mechanism. 
Note also that our $\psi'$ cross section in the CSM is rather 
large in comparison with the direct $J/\Psi$ cross section in 
the CSM. A smaller value which compares more favorably with 
the data could be obtained if one expressed the cross section 
in terms of quarkonium masses \cite{VAE95}. From the point of view 
presented here, this agreement appears coincidental since the cross 
sections are dominated by octet production.  

Perhaps the worst failure of 
the theory  is the $\chi_{c1}$ to $\chi_{c2}$ ratio 
in the feed-down contribution that has been measured in the WA11 experiment 
at $E=185\,$GeV \cite{LEM82}. We see that the prediction is far too small
 even after inclusion 
of color octet contributions. The low rate of $\chi_1$ production  is
due to the fact, as  already mentioned, 
that the gluon-gluon fusion channel is 
suppressed by $\alpha_s/\pi$ compared to $\chi_{c2}$ 
due to angular momentum constraints. Together with $J/\psi$ 
(and $\psi'$) polarization, discussed in Sect.~4, 
the failure to reproduce this ratio 
emphasizes the importance of yet other production mechanisms, presumably 
of higher twist, which are  naively suppressed by $\Lambda_{QCD}/m_c$ 
\cite{VAE95}.

\subsection{Pion-induced collisions}

\begin{figure}[t]
   \vspace{0cm}
   \epsfysize=8cm
   \epsfxsize=8cm
   \centerline{\epsffile{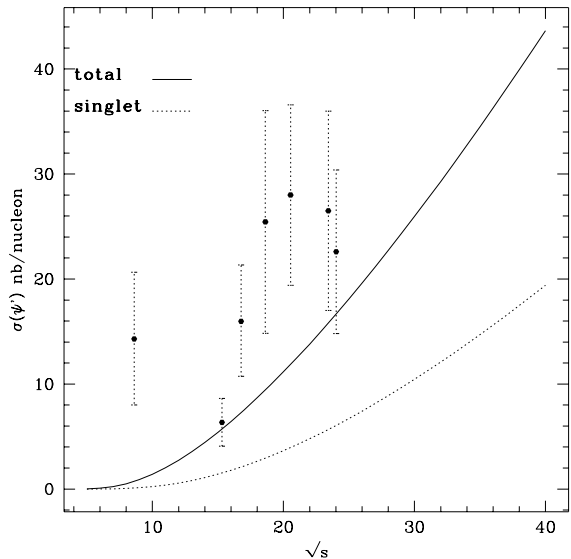}}
   \vspace*{0cm}
\caption{\label{pripifig} Total (solid) and singlet only (dotted) 
$\psi^\prime$ production cross section in pion-nucleon collisions 
($x_F>0$ only). The solid line is obtained 
with $\Delta_8(\psi')=5.2\cdot 10^{-3}\,$GeV${}^3$.} 
\end{figure} 
\begin{figure}[t]
   \vspace{0cm}
   \epsfysize=8cm
   \epsfxsize=8cm
   \centerline{\epsffile{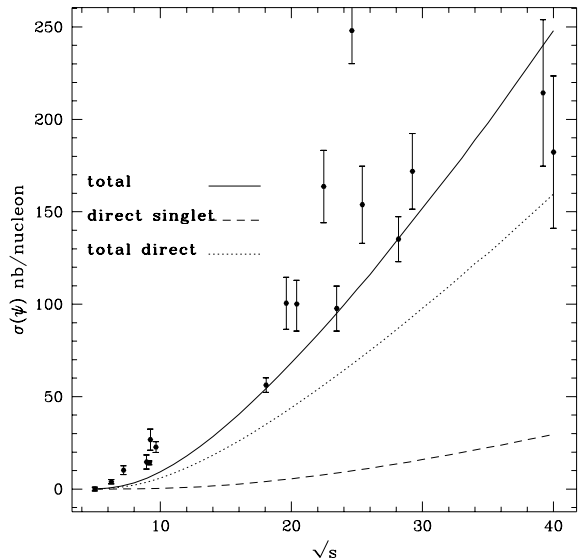}}
   \vspace*{0cm}
\caption{\label{psipifig} $J/\psi$ production cross sections in 
pion-nucleon collisions for $x_F>0$. Direct $J/\psi$ production 
in the CSM (dashed line) and after inclusion of color-octet processes 
(dotted line). The total cross section (solid line) includes radiative 
feed-down from the $\chi_{cJ}$ and $\psi'$ states. 
The solid line is obtained 
with $\Delta_8(J/\psi)=3.0\cdot 10^{-2}\,$GeV${}^3$.}
\end{figure}

The $\psi'$ and $J/\psi$ production cross section in 
pion-nucleon collisions are shown in 
Figs.~\ref{pripifig} and \ref{psipifig}. The discussion for 
proton-induced collisions applies with little modification to the pion 
case. A breakdown of contributions to the $J/\psi$ cross section 
at $E=300\,$GeV is given in table~\ref{tab3}. The theoretical 
prediction is based on the values of $\Delta_8(H)$ extracted from the 
proton data. Including color octet contributions can add 
little insight into the question 
of why the pion-induced cross sections appear to be 
systematically larger than expected. This issue has been 
extensively discussed in \cite{SCH94}. 
The discrepancy may be an indication that, either the 
gluon distribution in the pion is not really understood (although 
using parameterizations different from GRV LO tends to yield rather 
lower theoretical predictions), or that a genuine difference in 
higher twist effects for the proton and the pion exists. 

\subsection{$\Upsilon(nS)$}

\begin{figure}[t]
   \vspace{0cm}
   \epsfysize=8cm
   \epsfxsize=8cm
   \centerline{\epsffile{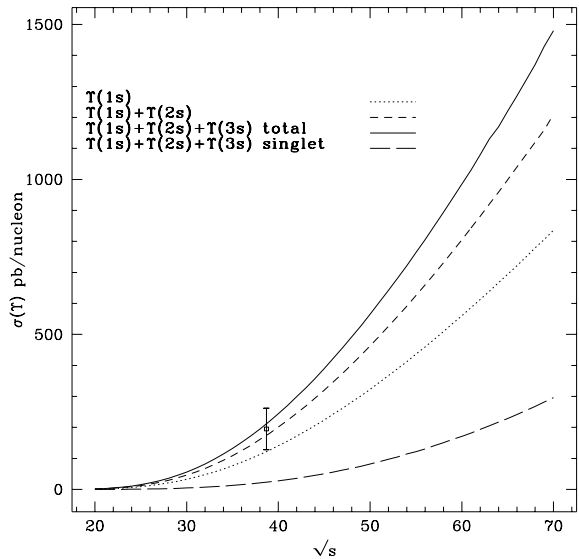}}
   \vspace*{0cm}
\caption{\label{upsfig} Total (direct plus indirect) $\Upsilon(nS)$ 
production cross sections (for $x_F>0$), 
consecutively summed over $n$. The data 
point refers to the sum of $n=1,2,3$. }
\end{figure}

If higher twist effects are important for fixed target charmonium 
production, their importance should decrease for bottomonium production 
and facilitate a test of color octet production. Unfortunately, 
data for bottomonium production at fixed target energies 
is sparse and does not allow us to 
complete this test.

{} Due to the 
increase of the quark mass, bottomonium production differs
in several ways from charmonium production,  from a theoretical 
standpoint. The relative 
quark-antiquark velocity squared decreases by a factor of three, 
thus, the color octet contributions to direct production of 
$\Upsilon(nS)$ are less important 
since they are suppressed by $v^4$ (at the same time $\alpha_s(2 m_Q)$ 
decreases much less). The situation is exactly the  opposite for 
the production of $P$-wave bottomonia.  In this case the color singlet 
and octet contributions scale equally in $v^2$. The increased quark mass,  
together with an increased relative importance of the octet 
matrix element $\langle {\cal O}_8^{\chi_{b0}} 
({}^3 S_1)\rangle$ (extracted from Tevatron data in \cite{CHO95II})
as compared to the singlet wavefunction (compare $\chi_{c0}$ with 
$\chi_{b0}$ in Tab.~\ref{tab2}), 
leads to domination of quark-antiquark pair initiated processes. 
Consequently, the direct 
$\Upsilon(nS)$ production cross section is at least a factor ten 
below the indirect contributions from $\chi_b$-decays. 
This observation leads to the 
conclusion that the number of  $\Upsilon(3S)$ observed by the 
E772 experiment \cite{ALD91} can only be explained if $\chi_{bJ} 
(3P)$ states that have not yet been observed directly exist below 
the open bottom threshold. Such indirect evidence has also been 
obtained from bottomonium production at the Tevatron collider 
\cite{PAP95}.  

To obtain our numerical results shown in Fig.~\ref{upsfig}, we 
assumed that these $\chi_{bJ}(3P)$ states decay into $\Upsilon(3S)$ 
with the same branching fractions as the corresponding $n=2$ states. 
The total cross sections are compared with the experimental value 
$195\pm 67\,$pb/nucleon obtained from \cite{ALE95} at $E=800\,$GeV 
for the sum of $\Upsilon(nS)$, $n=1,2,3$ and show very good agreement. 
The color-singlet processes alone would have led to a nine times 
smaller prediction at this energy. We should note, however, that 
integration of the $x_F$-distribution for $\Upsilon(1S)$ production 
given in \cite{ALD91} indicates a cross section about two to three times 
smaller than the central value quoted by \cite{ALE95}. The theoretical 
prediction for the relative production rates of $\Upsilon(1S):
\Upsilon(2S):\Upsilon(3S)$ is $1:0.42:0.30$ to be compared with 
the experimental ratio \cite{ALD91} $1:0.29:0.15$\footnote{
These numbers were taken from the raw data with no concern
regarding the differing efficiencies for the individual states.}. 
This comparison 
should not be over interpreted since it depends largely on the 
rather uncertain octet matrix elements for $P$-wave bottomonia. Due 
to lack of more data we also hesitate to use this comparison for 
a new determination of these matrix elements.

\section{$\psi'$ and $J/\psi$ Polarization}

In this section, we deal with $\psi'$ and $J/\psi$ polarization at 
fixed target energies and at colliders at large transverse momentum. 
Before returning to fixed target production in Sect.~4.2, we 
digress on large-$p_t$ production. We recall that, at large 
$p_t^2\gg 4 m_Q^2$, $\psi'$ and direct $J/\psi$ production is 
dominated by gluon fragmentation into color octet quark-antiquark pairs 
and expected to  yield  transversely polarized
quarkonia \cite{WIS95}. The reason for this is that a fragmenting gluon 
can be considered as on-shell and therefore transverse. Due to spin 
symmetry of NRQCD, the quarkonium inherits the transverse polarization 
up to corrections of order $4 m_c^2/p_T^2$ and $v^4$. Furthermore, 
it has been shown \cite{BEN95} that including radiative corrections to gluon 
fragmentation still leads to more than $90\%$ transversely polarized 
$\psi'$ (direct $J/\psi$). Thus, polarization provides one of the 
most significant tests for the color octet production 
mechanism at large transverse momenta. At moderate $p_t^2\sim 4 m_c^2$,  
non-fragmentation contributions proportional to 
$\langle {\cal O}_8^H ({}^1 S_0)\rangle$ and 
$\langle {\cal O}_8^H ({}^3 P_J)\rangle$ are sizeable \cite{CHO95II}. 
Understanding their polarization yield quantitatively is very 
important since most of the $p_t$-integrated data comes from the 
lower $p_t$-region. The calculation of the polarization yield 
has also been attempted in \cite{CHO95II}. However, the 
method used is at variance with \cite{BEN95} and leads to an 
incorrect result for $S$-wave quarkonia produced through 
intermediate quark-antiquark pairs in a color octet $P$-wave state. 
In the following subsection we expound on the method discussed in 
\cite{BEN95} and hope to clarify this difference.  

\subsection{Polarized production}

For arguments sake, let us consider the production of a $\psi'$ in 
a polarization state $\lambda$. This state can be reached through 
quark-antiquark pairs in various spin and orbital angular momentum states,  
and we are led to consider the intermediate quark-antiquark pair 
as a coherent superposition of these states. Because of parity 
and charge conjugation symmetry, intermediate states with 
different spin $S$ and angular momentum $L$ can not 
interfere\footnote{Technically, this means that NRQCD matrix elements 
with an odd number of derivatives or spin matrices vanish if the 
quarkonium is a $C$ or $P$ eigenstate.}, 
so that the only non-trivial situation occurs for ${}^3 P_J$-states, 
i.e. $S=1$, $L=1$.

In \cite{CHO95II} it is assumed that intermediate states with 
different $J J_z$, where $J$ is total angular momentum do not 
interfere, so that the production cross section can be expressed 
as the sum over $J J_z$ of the amplitude squared for production 
of a color octet quark-antiquark pair in a ${}^3 P_{J J_z}$ state 
times the amplitude squared for its transition into the $\psi'$. 
The second factor can be inferred from spin symmetry to be 
a simple Clebsch-Gordon coefficient so that 

\begin{equation}
\label{cho}
\sigma^{(\lambda)}_{\psi'} \sim \sum_{J J_z}
\sigma(\bar{c} c[{}^3 P^8_{J J_z}])\,|\langle J J_z|1 
(J_z-\lambda);1\lambda\rangle|^2\,.
\end{equation}

\noindent We will show that this equation is incompatible with spin 
symmetry which requires interference of intermediate 
states with different $J$.

A simple check can be obtained by applying (\ref{cho}) to 
the calculation of the gluon fragmentation function into 
longitudinally polarized $\psi'$. Since the fragmentation functions 
into quark-antiquark pairs in a ${}^3 P^8_{J J_z}$ state 
follow from \cite{TRI95} by a change of color factor, the 
sum in (\ref{cho}) can be computed. The result not only 
differs from the fragmentation function obtained in \cite{BEN95} 
but contains an infrared divergence which 
can not be absorbed into another NRQCD matrix element.

To see the failure of (\ref{cho}) more clearly we return to the 
NRQCD factorization formalism. After Fierz rearrangement of color 
and spin indices as explained in \cite{BOD95}, the cross 
section can be written as 

\begin{equation}
\label{factor}
\sigma^{(\lambda)} \sim H_{ai;bj}\cdot S_{ai;bj}^{(\lambda)}\,.
\end{equation}

\noindent In this equation $H_{ai;bj}$ is the hard scattering cross 
section, and $S_{ai;bj}$ is the soft (non-per\-tur\-ba\-tive) part that 
describes the `hadronization' of the color octet quark pair into 
a $\psi'$ plus light hadrons. Note that the 
statement of factorization entailed in this equation occurs only on 
the cross section and not on the amplitude level. The indices $ij$ and 
$ab$ refer to spin and angular momentum in a Cartesian basis $L_a S_i$ 
($a,i=1,2,3=x,y,z$). Since spin-orbit coupling is 
suppressed by $v^2$  in the NRQCD Lagrangian, $L_z$ and $S_z$ are good 
quantum numbers. In the specific situation we are considering,  
the soft part is simply given by (the notation follows 
\cite{BOD95,BEN95}) 

\begin{equation}
S_{ai;bj}^{(\lambda)}=
\langle 0|\chi^\dagger\sigma_i T^A\left(-\frac{i}{2} 
\stackrel{\leftrightarrow}{D}_a\right) 
\psi\,{a_{\psi'}^{(\lambda)}}^\dagger 
a_{\psi'}^{(\lambda)}\,\psi^\dagger\sigma_j T^A\left(-\frac{i}{2} 
\stackrel{\leftrightarrow}{D}_b\right)\chi|0\rangle\,,
\end{equation}

\noindent where $a_{\psi'}^{(\lambda)}$ destroys a $\psi'$ 
in an out-state with 
polarization $\lambda$. To evaluate this matrix element at leading order 
in $v^2$, we may use spin symmetry. Spin symmetry tells us that 
the spin of the $\psi'$ is aligned with the spin of the $\bar{c} c$ pair, 
so $S_{ai;bj}^{(\lambda)}\propto 
{\epsilon^i}^*(\lambda)\epsilon^j(\lambda)$. Now all 
vectors $S^{(\lambda)}_{ai;bj}$ 
can depend on have been utilized, and thus by rotational invariance, only the Kronecker 
symbol is left to tie up $a$ and $b$. The overall normalization is 
determined by taking appropriate contractions, and we obtain 

\begin{equation}
\label{decomp}
S_{ai;bj}^{(\lambda)}=
\langle {\cal O}^{\psi^\prime}_8(^3\!P_0)\rangle\,
\delta_{ab}\,{\epsilon^i}^*(\lambda)\epsilon^j(\lambda)\,.
\end{equation}

\noindent This decomposition tells us that to calculate 
the polarized production
rate we should project the hard scattering amplitude onto states
with definite $S_z=\lambda$ and $L_z$, square the 
amplitude, and then sum over 
$L_z$ ($\sum_{L_z}\epsilon_a(L_z)
\epsilon_b(L_z)=\delta_{ab}$  in the rest frame). In other words, the 
soft part is diagonal in the $L_z S_z$ basis.

It is straightforward to transform to the $J J_z$ basis. Since 
$J_z=L_z+S_z$, there is no interference between intermediate states 
with different $J_z$. To see this we write, in obvious 
notation, 

\begin{equation}
\label{factornew}
\sigma^{(\lambda)} \sim \sum_{J J_z;J' J_z^\prime} 
H_{J J_z;J' J_z^\prime}\cdot S_{J J_z;J' J_z^\prime}^{(\lambda)}\,,
\end{equation}

\noindent and  using (\ref{decomp}) obtain,  

\begin{equation}
S_{J J_z;J' J_z^\prime}^{(\lambda)} =
\langle {\cal O}_8^{\psi'} ({}^3 P_0)\rangle 
\sum_M \langle 1M;1\lambda|J J_z\rangle\langle J' J_z^\prime|1 M;1 \lambda
\rangle\,,
\end{equation}

\noindent which is diagonal in $(J J_z)(J' J_z^\prime)$ only after 
summation over $\lambda$ (unpolarized production). In general, the 
off-diagonal matrix elements cause interference of the following 
$J J_z$ states: $00$ with $20$, $11$ with $21$ and $1(-1)$ with $2 (-1)$. 
While the diagonal elements agree with (\ref{cho}), the off-diagonal 
ones are missed in (\ref{cho}).

To assess the degree of transverse $\psi'$ (direct $J/\psi$) 
polarization at moderate $p_t$, 
the calculation of \cite{CHO95II} should be redone with the correct 
angular momentum projections. 

\subsection{Polarization in fixed target experiments}

Polarization measurements have been performed for both $\psi$ 
\cite{AKE93} and 
$\psi^\prime$ \cite{HEI91} production 
in pion scattering fixed target experiments. 
Both experiments observe an essentially flat angular distribution in 
the decay $\psi\to \mu^+ \mu^-$ ($\psi= J/\psi,\psi'$), 

\begin{equation}
\frac{d\sigma}{d\cos\theta }\propto 1+ \alpha \cos^2 \theta\,,
\end{equation}

\noindent where the angle $\theta$ is defined as the angle between 
the three-momentum vector of the positively charged muon and 
the beam axis in the rest frame of the quarkonium. The observed values 
for $\alpha$ are $0.02\pm 0.14$ for $\psi'$, measured at 
$\sqrt{s}=21.8\,$GeV in the region $x_F>0.25$ and 
$0.028\pm 0.004$ for $J/\psi$ measured at $\sqrt{s}=15.3\,$GeV 
in the region $x_F>0$. In the CSM, the $J/\psi$'s are predicted to be 
significantly transversely polarized \cite{VAE95}, in conflict with 
experiment.

The polarization yield of color octet processes can be calculated 
along the lines of the previous subsection. We first concentrate on 
$\psi'$ production and define $\xi$ as the fraction of longitudinally 
polarized $\psi'$. It is related to $\alpha$ by 

\begin{equation}
\alpha=\frac{1-3\xi}{1+\xi}\,.
\end{equation}

\noindent For the different intermediate quark-antiquark 
states we find the following ratios of longitudinal to transverse 
quarkonia:

\begin{equation}
\addtolength{\arraycolsep}{0.3cm}
\begin{array}{ccc}
{}^3 S_1^{(1)} & 1:3.35 & \xi=0.23\\
{}^1 S_0^{(8)} & 1:2 & \xi=1/3\\
{}^3 P_J^{(8)} & 1:6 & \xi=1/7\\
{}^3 S_1^{(8)} & 0:1 & \xi=0
\end{array}
\end{equation}

\noindent where the number for the singlet process (first line) has been 
taken from \cite{VAE95}\footnote{This number is $x_F$-dependent and we 
have approximated it 
by a constant at low $x_F$, where the bulk data is obtained from. 
The polarization fractions for the octet $2\to 2$ parton processes 
are $x_F$-independent.}. Let us add the following remarks: 

(i) The ${}^3 S_1^{(8)}$-subprocess yields pure transverse 
polarization. Its contribution to the total polarization is not 
large, because gluon-gluon fusion dominates the total rate.

(ii) For the ${}^3 P_J^{(8)}$-subprocess $J$ is not specified, 
because interference between intermediate states with different 
$J$ could occur as discussed in the previous subsection. As it 
turns out, interference does in fact not occur at leading order 
in $\alpha_s$, because the only non-vanishing short-distance 
amplitudes in the $J J_z$ basis are $00$, $22$ and 
$2(-2)$, which do not 
interfere.

(iii) The ${}^1 S_0^{(8)}$-subprocess yields unpolarized quarkonia. 
This follows from the fact that the NRQCD matrix element is 

\begin{equation}
\label{above}
\langle 0|\chi^\dagger T^A{a_{\psi'}^{(\lambda)}}^\dagger 
a_{\psi'}^{(\lambda)}\,\psi^\dagger T^A\chi|0\rangle =\frac{1}{3}\,
\langle {\cal O}_8^{\psi'}({}^1 S_0)\rangle\,,
\end{equation}

\noindent independent of the helicity state $\lambda$. At this point, 
we differ from \cite{TAN95}, who assume that this channel results 
in pure transverse polarization, because the gluon in the 
chromomagnetic dipole transition ${}^1 S_0^{(8)}\to {}^3 S_1^{(8)}+g$ 
is assumed to be transverse. However, one should keep in mind that 
the soft gluon is off-shell and interacts with other partons with unit 
probability  prior to  hadronization. The NRQCD formalism 
applies only to inclusive quarkonium production. Eq.~(\ref{above}) 
then follows from rotational invariance.

(iv) Since the ${}^3 P_J^{(8)}$ and ${}^1 S_0^{(8)}$-subprocesses 
give different longitudinal polarization fractions, the $\psi'$ 
polarization depends on a combination of the matrix elements 
$\langle {\cal O}_8^{\psi'} ({}^1 S_0)\rangle$ and 
$\langle {\cal O}_8^{\psi'} ({}^3 P_0)\rangle$ which is different 
from $\Delta_8(\psi')$. 

To obtain the total polarization the various subprocesses have to be 
weighted by their partial cross sections. We define 

\begin{equation}
\delta_8(H)=\frac{\langle {\cal O}_8^{H} ({}^1 S_0)\rangle}
{\Delta_8(H)}
\end{equation}

\noindent and obtain 

\begin{eqnarray}
\xi &=& 0.23\,\frac{\sigma_{\psi'}({}^3 S_1^{(1)})}{\sigma_{\psi'}} + 
\left[\frac{1}{3}\delta_8(\psi')+\frac{1}{7} (1-\delta_8(\psi'))\right] 
\frac{\sigma_{\psi'}({}^1 S_0^{(8)}+{}^3 P_J^{(8)})}{\sigma_{\psi'}}
\nonumber\\
&=& 0.16+0.11\,\delta_8(\psi')\,,
\end{eqnarray}

\noindent where the last line holds at $\sqrt{s}=21.8\,$GeV (The 
energy dependence is mild and the above formula can be used with 
little error even at $\sqrt{s}=40\,$GeV). Since $0<\delta_8(H)<1$, we
have $0.16<\xi<0.27$ and therefore

\begin{equation}
0.15 < \alpha < 0.44\,.
\end{equation}

\noindent In quoting this range we do not attempt an estimate of 
$\delta_8(\psi')$. Note that taking the Tevatron and fixed target 
extractions of certain (and different) combinations of 
$\langle {\cal O}_8^{\psi'} ({}^1 S_0)\rangle$ and 
$\langle {\cal O}_8^{\psi'} ({}^3 P_0)\rangle$ seriously 
(see Sect.~5.1), a large value of $\delta_8(\psi')$ 
and therefore low $\alpha$ would be favored. Within large errors, 
such a scenario could be considered consistent with the 
measurement quoted earlier. From a theoretical point of view, however, 
the numerical violation of velocity counting rules implied by 
this scenario would be rather disturbing.

In contrast, the more accurate measurement of polarization for 
$J/\psi$ leads to a clear discrepancy with theory. In this case, we
have to incorporate the polarization inherited from decays of 
the higher charmonium states $\chi_{cJ}$ and $\psi'$. This task 
is simplified by observing that the 
contribution from $\chi_{c0}$ and $\chi_{c1}$ 
feed-down is (theoretically) small as is the octet contribution 
to the $\chi_{c2}$ production cross section. On the other hand, the 
gluon-gluon fusion process produces $\chi_{c2}$ states only in 
a helicity $\pm 2$ level, so that the $J/\psi$ in the subsequent 
radiative decay is completely transversely polarized. 
Weighting all subprocesses by their partial cross section 
and neglecting the small $\psi'$ feed-down, we arrive at 

\begin{equation}
\xi = 0.10 + 0.11\,\delta_8(J/\psi)
\end{equation}

\noindent at $\sqrt{s}=15.3\,$GeV, again with mild energy dependence.
This translates into sizeable transverse polarization

\begin{equation}
0.31 < \alpha < 0.63\,.
\end{equation}

\noindent The discrepancy with data could be ameliorated if the observed
number of $\chi_{c1}$ from 
feed-down were used instead of the theoretical value. However, 
we do not know the polarization yield of whatever mechanism is 
responsible for copious $\chi_{c1}$ production.

Thus, color octet mechanisms do not help to solve the 
polarization problem and 
one has to invoke a significant higher-twist contribution as 
discussed in \cite{VAE95}. To our knowledge, no specific mechanism 
has yet been proposed that would yield predominantly longitudinally 
polarized $\psi'$ and $J/\psi$ in the low $x_F$ region which dominates 
the total production cross section. One might speculate that both 
the low $\chi_{c1}/\chi_{c2}$ ratio and the large transverse polarization
follow from the assumption of transverse gluons in the gluon-gluon fusion 
process, as inherent to the leading-twist approximation. If gluons 
in the proton and pion have 
large intrinsic transverse momentum, as suggested by the 
$p_t$-spectrum in open charm 
production, one would be naturally led to higher-twist effects that 
obviate the helicity constraint on on-shell gluons.

\section{Other processes}

Direct $J/\psi$ and $\psi'$ production is sensitive to the color octet 
matrix element $\Delta_8(H)$ defined in (\ref{delta}). In this section we 
compare our extraction of $\Delta_8(H)$ with constraints from 
quarkonium production at the Tevatron and in photo-production at fixed 
target experiments and HERA.

\subsection{Quarkonium production at large $p_t$}

An extensive analysis of charmonium production data at $p_t>5\,$GeV 
has been carried out by Cho and Leibovich \cite{CHO95,CHO95II}, who 
relaxed the fragmentation approximation employed earlier 
\cite{BRA95,CAC95}. At the lower $p_t$ boundary, the theoretical 
prediction is dominated by the ${}^1 S_0^{(8)}$ and ${}^3 P_J^{(8)}$ 
subprocesses and the fit yields

\begin{eqnarray}
\label{tevme}
\langle {\cal O}^{J/\psi}_8 ({}^1 S_0) \rangle + 
\frac{3}{m_c^2}\langle {\cal O}^{J/\psi}_8 ({}^3 P_0) \rangle\,
= 6.6\cdot 10^{-2}\nonumber\\
\langle {\cal O}^{\psi'}_8 ({}^1 S_0) \rangle + 
\frac{3}{m_c^2}\langle {\cal O}^{\psi'}_8 ({}^3 P_0) \rangle\,
= 1.8\cdot 10^{-2}\,,
\end{eqnarray}

\noindent to be compared with the fixed target values\footnote{
Since there is a strong correlation between the charm quark mass 
and the extracted NRQCD matrix elements, we emphasize that both 
(\ref{tevme}) and (\ref{fixme}) as well as (\ref{photome}) below 
have been obtained with the same $m_c=1.5\,$GeV  (or $m_c=1.48\,$ GeV, 
to be precise). On the other hand, the apparent agreement of 
predictions for fixed target experiments with data claimed in 
\cite{GUP96} is obtained from (\ref{photome}) in conjunction with 
$m_c=1.7\,$GeV.}

\begin{eqnarray}\label{fixme}
\langle {\cal O}^{J/\psi}_8 ({}^1 S_0) \rangle + 
\frac{7}{m_c^2}\langle {\cal O}^{J/\psi}_8 ({}^3 P_0) \rangle\,
= 3.0\cdot 10^{-2}\nonumber\\
\langle {\cal O}^{\psi'}_8 ({}^1 S_0) \rangle + 
\frac{7}{m_c^2}\langle {\cal O}^{\psi'}_8 ({}^3 P_0) \rangle\,
= 0.5\cdot 10^{-2}\,.
\end{eqnarray}

\noindent If we assume $\langle {\cal O}^{J/\psi}_8 ({}^1 S_0) \rangle =
\langle {\cal O}^{J/\psi}_8 ({}^3 P_0) \rangle/m_c^2$, the 
fixed target values are a factor seven (four) smaller than the 
Tevatron values for $J/\psi$ ($\psi'$). The discrepancy would be lower 
for the radical choice $\langle {\cal O}^{J/\psi}_8 ({}^3 P_0) \rangle=0$. 

While this comparison looks like a flagrant violation of the 
supposed process-independence of NRQCD production matrix 
elements, there are at least two possibilities that could lead to 
systematic differences:

 (i) The $2\to 2$ color octet parton processes are 
schematically of the form

\begin{equation}
\frac{\langle {\cal O}\rangle}{2 m_c}\,\frac{1}{M_f^2}\,
\delta(x_1 x_2 s-M_f^2)\,,
\end{equation}

\noindent where $M_f$ denotes the final state invariant mass. To leading 
order in $v^2$, we have $M_f=2 m_c$. Note, however, that this is 
physically unrealistic. Since color must be emitted from the quark pair 
in the octet state and neutralized by final-state interactions, the 
final state is a quarkonium accompanied by light hadrons with 
invariant mass squared of order $M_f^2\approx (M_H+M_H v^2)^2$ 
since the soft gluon emission carries an energy of order $M_H v^2$, 
where $M_H$ is the quarkonium mass. The kinematic effect of this 
difference in invariant mass is very large since the gluon 
distribution rises steeply at small $x$ and reduces the 
cross section by at least a factor two. 
The `true' matrix elements would therefore be larger 
than those extracted from fixed target experiments at leading order 
in NRQCD. Since the $\psi'$ is heavier than the $J/\psi$, the effect 
is more pronounced for $\psi'$, consistent with the larger 
disagreement with the Tevatron extraction for $\psi'$. Note that 
the effect is absent for large-$p_t$ production, since in this 
case,  $x_1 x_2 s > 4 p_t^2 \gg M_f^2$. If we write $M_f=2 m_c+{\cal O} 
(v^2)$,  then the difference between fixed target and large-$p_t$ production 
stems from  different behaviors of the velocity expansion in the two cases.

(ii) It is known that small-$x$ effects increase the 
open bottom production cross section at the Tevatron as compared 
to collisions at lower $\sqrt{s}$. Since even at large $p_t$, 
the typical $x$ is smaller 
at the Tevatron than in fixed target experiments, this effect would 
enhance the Tevatron prediction more than the fixed target prediction. 
The `true' matrix elements would therefore be smaller than 
those extracted from the Tevatron in \cite{CHO95II}. 

While a combination of both effects could well account for the 
apparently different NRQCD matrix elements, one must keep in mind 
that we have reason to suspect important higher twist effects 
for charmonium production at fixed target energies. Theoretical 
predictions for fixed target production are intrinsically less 
accurate than at large $p_t$, where higher-twist contributions 
due to the initial hadrons are expected to be suppressed by 
$\Lambda_{QCD}/p_t$ (if not $\Lambda_{QCD}^2/p_t^2$) 
rather than $\Lambda_{QCD}/m_c$.

\subsection{Photo-production}

A comparison of photo-production with fixed target production is 
more direct since the same combination of NRQCD matrix elements 
is probed and the kinematics is similar. All analyses 
\cite{CAC96,AMU96,KO96} find a substantial overestimate of the 
cross section if the octet matrix elements of (\ref{tevme}) are used. 
The authors of \cite{AMU96} fit

\begin{equation}\label{photome}
\langle {\cal O}^{J/\psi}_8 ({}^1 S_0) \rangle + 
\frac{7}{m_c^2}\langle {\cal O}^{J/\psi}_8 ({}^3 P_0) \rangle\,
= 2.0\cdot 10^{-2}\,,
\end{equation}

\noindent consistent with (\ref{fixme}) within errors, which we have not 
specified. While this agreement is 
reassuring, it might also be partly accidental since the extraction 
of \cite{AMU96} is performed on the elastic peak, which is not 
described by NRQCD. Color octet mechanisms do not 
leave a clear signature in the total inelastic photo-production cross 
section. The authors of \cite{CAC96} argue that the color-octet 
contributions to the energy spectrum of $J/\psi$ are in conflict 
with the observed energy dependence in the endpoint region 
$z>0.7$, where $z=E_{J/\psi}/E_\gamma$ in the proton rest frame. 
This discrepancy would largely disappear if the smaller matrix 
element of (\ref{fixme}) or (\ref{photome}) were used rather than 
(\ref{tevme}). Furthermore, since in a color octet process soft 
gluons with energy $M_H v^2$ must be emitted, but are kinematically 
not accounted for, the NRQCD-prediction for the energy distribution 
should be smeared over an interval of size $\delta z\sim v^2\sim 0.3$, 
making the steep rise of the energy distribution close to $z=1$ is 
not necessarily physical.

\section{Conclusion}

We have reanalyzed charmonium production data from fixed target 
experiments, including color octet production mechanisms. 
Our conclusion is twofold: On one hand, the inclusion of 
color octet processes allows us to reproduce the overall normalization 
of the total production cross section with color octet matrix 
elements of the expected size (if not somewhat smaller) 
without having to invoke small values of the charm quark mass. 
This was found to be true for bottomonium as well as for charmonium.
Comparing the theoretical 
predictions within this framework with the data 
implies the existence 
of additional bottomonium states below threshold which have not yet 
been seen directly.

On the other hand, the present picture of charmonium production at 
fixed target energies is far from perfect. The $\chi_{c1}/
\chi_{c2}$ production ratio remains almost an order of magnitude 
too low, and the transverse polarization fraction  
of the  $J/\psi$ and $\psi'$ is too large. We thus confirm the expectation of 
\cite{VAE95} 
that higher twist effects must be substantial even after including
the octet mechanism.

The uncertainties in the theoretical prediction at fixed target 
energies are substantial and preclude a straightforward test 
of universality of color octet matrix elements by comparison with 
quarkonium production at large transverse momentum. We have 
argued that small-$x$, as well as kinematic effects, could bias 
the extraction of these matrix elements in different directions 
at fixed target and collider energies. The large uncertainties 
involved, especially due to the charm quark mass, could hardly 
be eliminated by 
a laborious calculation of $\alpha_s$-corrections to the 
production processes considered here. 
To more firmly establish existence of 
the octet mechanism there are several experimental
measurements which need to be performed. Data on polarization is presently
only available for charmonium production in pion-induced collisions.
A measurement of polarization at large transverse momentum or 
for bottomonium is of crucial importance, 
because higher twist effects should be suppressed. 
Furthermore, a measurement of direct
and indirect production fractions in the bottom system would provide further 
confirmation of the color octet picture and constrain the color octet 
matrix elements for bottomonium. 

\vspace*{1cm}
\noindent {\bf Acknowledgments.} We thank S.J.~Brodsky, 
E.~Quack  and V.~Sharma for discussions. IZR acknowledges support from the
DOE grant DE-FG03-90ER40546 and the NSF grant PHY-8958081.

\vfill\eject

\end{document}